\documentstyle[aps,preprint,eqsecnum,floats,epsf]{revtex}
\tighten
\draft
\begin{document}

\preprint{MAD-NT-96-02}
\title{Semiclassical treatment of matter-enhanced neutrino
oscillations for an arbitrary density profile}
\author{A.~B. Balantekin\thanks{Electronic address: {\tt
	baha@nucth.physics.wisc.edu}}
	and
	J.~F. Beacom\thanks{Electronic address: {\tt
	beacom@nucth.physics.wisc.edu}}}
\address{Department of Physics, University of Wisconsin\\
        Madison, WI 53706}
\date{\today}
\maketitle

\begin{abstract}
The matter-enhanced oscillations of two neutrino flavors are studied
using a uniform semiclassical approximation.  Unlike some analytic
studies which have focused on certain exactly-solvable densities, this
method can be used for an arbitrary monotonic density profile.  The
method is applicable to a wider range of mixing parameters than
previous approximate methods for arbitrary densities.  The
approximation is excellent in the adiabatic regime and up to the
extreme nonadiabatic limit.  In particular, the range of validity for
this approximation extends farther into the nonadiabatic regime than
for the linear Landau-Zener result.  This method also allows
calculation of the source- and detector-dependent terms in the
unaveraged survival probability, and analytic results for these terms
are given.  These interference terms may be important in studying
neutrino mixing in the sun or in supernovae.
\end{abstract}



\newpage
\section{Introduction}

Matter-enhanced oscillations of neutrino flavors via the
Mikheyev-Smirnov-Wolfenstein (MSW) mechanism \cite{MSW} have been
studied for neutrinos in various environments, but most extensively
for the sun, in connection with the solar neutrino problem
\cite{Bahcall}.  For a recent review of the solar neutrino problem and
the ongoing neutrino detection experiments, see
Ref.~\cite{Haxton-review}.  Recently, interest has also been
developing for the study of neutrino oscillations in supernovae
\cite{Snova}.

The approximate results derived in this paper are applicable to
matter-enhanced two-flavor neutrino oscillations in general physical
situations.  Analytic results are important for several reasons.
While numerical integration of the MSW equations is straightforward,
it becomes extremely tedious when it must be done for a large range of
the mixing parameters.  Analytic results also allow a greater
understanding of the effects of changes in the parameters, and may be
useful for extracting information about the solar density from the
measured neutrino fluxes.

Analytic studies of matter-enhanced neutrino oscillations proceed
along two lines.  The first approach is the study of certain densities
for which an exact solution for the oscillation probability can be
obtained.  The mixing parameters are allowed to be arbitrary.  The
exponential density has attracted particular interest, since it
approximates the solar density.  A catalog of all of the exactly
solvable densities has been presented in Ref.~\cite{Bychuk}.  The
second approach allows for a general density, but restricts the
parameters so that an approximation can be made to the equations of
motion, which are then solved exactly.  The approximations are chosen
so that the exact results are recovered in either the extreme
nonadiabatic or extreme adiabatic limit.  In this paper, we consider a
uniform semiclassical approximation to derive the neutrino conversion
probability for an arbitrary density.  The solution is exact in the
adiabatic limit, like the linear Landau-Zener result.  However, the
new result has a larger range of validity in the nonadiabatic regime.
In the body of the paper, we will discuss how some of the different
approximations are related.


\section{Matter-Enhanced Neutrino Oscillations}


\subsection{Coupled Equations in the Flavor Basis}

For two neutrino flavors (taken here to be electron and muon) in
matter, the equations of motion for the $\nu_e$ and $\nu_{\mu}$
probability amplitudes in the relativistic limit are
\begin{equation}
i\hbar \frac{\partial}{\partial t}
\left[\begin{array}{cc} \Psi_e(t) \\ \\ \Psi_{\mu}(t)\end{array}\right]
=
\frac{1}{4 E}
\left[\begin{array}{cc}
A - \delta m^2 \cos{2\theta_v} & \delta m^2 \sin{2\theta_v}\\ \\
\delta m^2 \sin{2\theta_v} & -A + \delta m^2 \cos{\theta_v}
\end{array}\right]
\left[\begin{array}{cc} \Psi_e(t) \\ \\ \Psi_{\mu}(t) \end{array}\right]\,,
\end{equation}
where all terms in the Hamiltonian proportional to the identity have
been dropped since they do not contribute to the relative phase
between the $\nu_e$ and $\nu_{\mu}$ components.  The vacuum mixing
parameters are specified by the vacuum mixing angle $\theta_v$, taken
to be $0 < \theta_v < \pi/4$, and the vacuum mass-squared splitting
$\delta m^2 \equiv m_2^2 - m_1^2$, where we take $m_2 > m_1$.
Electron neutrinos experience charged-current scattering with the
electrons in the medium, whereas muon neutrinos do not.  This
difference yields the effective mass correction
\begin{equation}
A = 2 \sqrt{2}\ G_F N_e(t) E\,,
\end{equation}
where $G_F$ is the Fermi constant and $N_e(t)$ is the number density
of electrons in the medium.

Before proceeding further, we switch to working with dimensionless
quantities.  We define a length scale
\begin{equation}
L = \frac{\hbar\lambda}{\delta m^2 /4 E}\,,
\label{eq:L}
\end{equation}
and use this to define $x = t/L$.  Since we will be making a
semiclassical expansion, we need to be able to keep track of formal
powers of $\hbar$.  For each $\hbar$ in the problem, we write
$\lambda$ and consider $\lambda$ to be formally small; this is
equivalent to saying that the length $L$ is small.  We will make
expansions in powers of $\lambda$, truncating the higher orders.  At
the end of the calculation, we will set $\lambda = 1$.  For notational
convenience, we write
\begin{equation}
i\lambda \frac{\partial}{\partial x}
\left[\begin{array}{cc} \Psi_e(x) \\ \\ \Psi_{\mu}(x) \end{array}\right]
= H_{e\mu}(x)
\left[\begin{array}{cc} \Psi_e(x) \\ \\ \Psi_{\mu}(x) \end{array}\right]
=
\left[\begin{array}{cc}
\eta\varphi(x) & \sqrt{\Lambda} \\ \\
\sqrt{\Lambda} & -\eta\varphi(x)
\end{array}\right]
\left[\begin{array}{cc} \Psi_e(x) \\ \\ \Psi_{\mu}(x) \end{array}\right]\,.
\label{eq:Hflav}
\end{equation}
We have defined
\begin{equation}
\eta\varphi(x) = \zeta(x) - \cos{2\theta_v}
\label{eq:varphi}
\end{equation}
and
\begin{equation}
\Lambda = \sin{2\theta_v}\,.
\label{eq:Lambda}
\end{equation}
The scaled electron density is
\begin{equation}
\zeta(x) = \zeta(x_i)N_e(x)/N_e(x_i)\,,
\end{equation}
normalized at the initial point $x_i$ as
\begin{equation}
\zeta(x_i) = \frac{2 \sqrt{2}\ G_F E N_e(x_i)}{\delta m^2}\,.
\end{equation}
Note that there are notation changes from
Ref.~\cite{Susyform1,Susyform2,Susyform3}; here we have made $\Lambda$
and $\varphi$ dimensionless.  The factor $\eta$ (taken to be $\pm 1$),
is introduced above to control the analytic behavior of the function
$\varphi(x)$ in the complex plane, as explained in the Appendix.  In
the expressions with $\varphi^2$ below, we drop $\eta^2 = 1$.


\subsection{Coupled Equations in the Adiabatic Basis}
\label{sec:adi}

The flavor-basis Hamiltonian of Eq.~(\ref{eq:Hflav}) can be
instantaneously diagonalized.  We make a change of basis
\begin{equation}
\left[\begin{array}{cc} \Psi_1(x) \\ \\ \Psi_2(x) \end{array}\right]
= R(-\theta(x))
\left[\begin{array}{cc} \Psi_e(x) \\ \\ \Psi_{\mu}(x) \end{array}\right]
=
\left[\begin{array}{cc}
\cos{\theta(x)} & -\sin{\theta(x)} \\ \\
\sin{\theta(x)} & \cos{\theta(x)}
\end{array}\right]
\left[\begin{array}{cc} \Psi_e(x) \\ \\ \Psi_{\mu}(x) \end{array}\right]\,.
\end{equation}
$\Psi_1(x)$ is the probability amplitude to be in the ``light''
(primarily electron type) eigenstate in the mass basis, and
$\Psi_2(x)$ is the probability amplitude to be in the ``heavy''
(primarily muon type) eigenstate in the mass basis.  The requirement
that this transformation instantaneously diagonalize $H_{e\mu}(x)$
defines the matter mixing angle via
\begin{equation}
\sin{2\theta(x)} = \frac{\sqrt{\Lambda}}{\sqrt{\Lambda + \varphi^2(x)}}
\label{eq:s2m}
\end{equation}
and
\begin{equation}
\cos{2\theta(x)} =  \frac{-\eta\varphi(x)}{\sqrt{\Lambda + \varphi^2(x)}}\,.
\label{eq:c2m}
\end{equation}
The matter angle thus ranges from $\pi/2$ at infinite density to
$\theta_v$ in vacuum.  At the resonance, $\theta = \pi/4$.  The
instantaneous eigenvalues of $H_{e\mu}(x)$ are
\begin{equation}
\mp \sqrt{\Lambda + \varphi^2(x)}\,,
\end{equation}
corresponding to $\Psi_1$ (the ``light'' eigenstate) and $\Psi_2$ (the
``heavy'' eigenstate), respectively.  The splitting between the
instantaneous mass eigenstates has a minimum as a function of $x$ when
$\varphi(x) = 0$, or $\zeta(x) = \cos{2\theta_v}$; this is the MSW
resonance point, which will be denoted by $x_c$.  The trajectories of
these eigenvalues represent an avoided level crossing.  The adiabatic
limit is the case where the neutrino stays in one of the instantaneous
eigenstates during its entire propagation.  In the nonadiabatic limit,
the neutrino may ``hop'' from one eigenstate to the other near the
resonance.

In the mass basis, the equations of motion are
\begin{equation}
i\lambda \frac{\partial}{\partial x}
\left[\begin{array}{cc} \Psi_1(x) \\ \\ \Psi_2(x) \end{array}\right]
= H_{12}(x)
\left[\begin{array}{cc} \Psi_1(x) \\ \\ \Psi_2(x) \end{array}\right]
=
\left[\begin{array}{cc}
-\sqrt{\Lambda + \varphi^2(x)} & -i \lambda \theta'(x) \\ \\
i \lambda \theta'(x) & \sqrt{\Lambda + \varphi^2(x)}
\end{array}\right]
\left[\begin{array}{cc} \Psi_1(x) \\ \\ \Psi_2(x) \end{array}\right]\,.
\label{eq:Hadi}
\end{equation}
Throughout the paper, prime denotes derivative with respect to x.
When the density is changing slowly, then so is the matter angle
$\theta(x)$, and the off-diagonal terms can be neglected; for that
reason, this is also known as the ``adiabatic'' basis.  The
adiabaticity parameter is defined as
\begin{equation}
\gamma(x) =
\left|\frac{\sqrt{\Lambda + \varphi^2(x)}}{i \lambda \theta'(x)}\right|\,,
\end{equation}
where $\theta'(x)$ can be derived from Eqs.~(\ref{eq:s2m}) and
(\ref{eq:c2m}).  When this parameter $\gamma(x)$ is large, we can
neglect the off-diagonal terms.  All nonadiabatic behavior, i.e.,
hopping from one mass eigenstate to the other, takes place in a
neighborhood of the resonance.  It is there that $\gamma(x)$ is
minimized, so the requirement of $\gamma(x) \gg 1$ for adiabatic
propagation is the most exacting:
\begin{equation}
\gamma_c = \gamma(x_c) =
\frac{2}{\lambda}
\frac{\sin^2{2\theta_v}}{\cos{2\theta_v}}
\frac{1}{\;\left|\zeta'/\zeta\right|_{x_c}}
\gg 1\,.
\end{equation}
In this limit, the equations of motion can be integrated immediately,
yielding pure phases for $\Psi_1(x)$ and $\Psi_2(x)$.  At the initial
point $x_i$, we take the neutrino to be a pure $\nu_e$, so
\begin{equation}
\Psi_e(x_i) = 1
\end{equation}
and
\begin{equation}
\Psi_e'(x_i) =
\left.\frac{\partial \Psi_e(x)}{\partial x}\right|_{x_i} =
-\frac{i}{\lambda} \eta\varphi(x_i)\,.
\end{equation}
The latter follows from $\Psi_e(x_i) = 1$, $\Psi_{\mu}(x_i) = 0$.  We
denote the initial matter angle by $\theta_i$ and reference the phases
from the initial point $x_i$.  The phase integral will be denoted by
\begin{equation}
I_p(x,x_i) \equiv
\int^x_{x_i} {dx \sqrt{\Lambda + \varphi^2(x)}}\,.
\end{equation}
Taking into account the basis changes at the initial and final points,
the adiabatic solutions, valid in the limit $\gamma(x_c) \gg 1$, are
\begin{equation}
\Psi_e(x) =
\cos{\theta(x)}\cos{\theta_i}
\exp\left(+\frac{i}{\lambda}I_p(x,x_i)\right) +
\sin{\theta(x)}\sin{\theta_i}
\exp\left(-\frac{i}{\lambda}I_p(x,x_i)\right)
\end{equation}
and
\begin{equation}
\Psi_{\mu}(x) =
-\sin{\theta(x)}\cos{\theta_i}
\exp\left(+\frac{i}{\lambda}I_p(x,x_i)\right) +
\cos{\theta(x)}\sin{\theta_i}
\exp\left(-\frac{i}{\lambda}I_p(x,x_i)\right)\,.
\end{equation}
These forms hold both before and after the resonance in the adiabatic
limit.  If nonadiabatic corrections are taken into account, then the
wave functions will have these forms before the resonance but will be
more complicated after the resonance.  In the adiabatic limit
\cite{Barger}, the electron neutrino survival probability at a general
point $x$ is:
\begin{eqnarray}
P(\nu_e \rightarrow \nu_e)(x,x_i) & = & \left|\Psi_e(x)\right|^2
\nonumber \\
& = & \frac{1}{2} \left[1 + \cos{2\theta_i}\cos{2\theta(x)}\right]
+ \frac{1}{2} \sin{2\theta_i}\sin{2\theta(x)}
\cos\left(\frac{2}{\lambda}I_p(x,x_i)\right)\,.
\end{eqnarray}
Note that the second term depends upon the source and detector
positions, and will disappear under averaging of either.  The
probability of conversion to muon type is given by $P(\nu_e
\rightarrow \nu_{\mu}) = 1 - P(\nu_e \rightarrow \nu_e)$.  If the
final point is chosen in vacuum, then $\theta(x) \rightarrow
\theta_v$.

With the adiabatic limit in hand, the obvious thing to do is to seek
the corrections that take into account $P_{hop}$, the probability of
hopping from one mass eigenstate to the other.  Above, the adiabatic
approximation was controlled by the ratio of diagonal to off-diagonal
elements.  That ratio is in turn controlled by $\lambda$, which keeps
track of powers of $\hbar$.  In the semiclassical limit of $\hbar
\rightarrow 0$, one has $\lambda \rightarrow 0$ and $\gamma_c
\rightarrow \infty$.  Note that $\lambda$ appears above in the
adiabatic survival probability only in the phase; the fully averaged
expression is independent of $\lambda$.  The way to treat $P_{hop}$
systematically is to expand in powers of $\lambda$ and to keep only
the lowest-order terms.


\subsection{Uncoupled Equations in the Flavor Basis}

The coupled first-order equations for the flavor-basis wave functions
can be decoupled to yield
\begin{equation}
-\lambda^2\frac{\partial^2 \Psi_e(x)}{\partial x^2} -
\left[\Lambda + \varphi^2(x) + i\lambda\eta\varphi'(x)\right]\Psi_e(x) = 0
\label{eq:SE_e}
\end{equation}
and
\begin{equation}
-\lambda^2\frac{\partial^2 \Psi_{\mu}(x)}{\partial x^2} -
\left[\Lambda + \varphi^2(x) - i\lambda\eta\varphi'(x)\right]\Psi_{\mu}(x) = 0
\,,
\label{eq:SE_m}
\end{equation}
where $\varphi(x)$ and $\Lambda$ are defined in Eqs.~(\ref{eq:varphi})
and (\ref{eq:Lambda}).  Such a simple decoupling is not possible in
the matter basis.

These Schr\"odinger-like equations are similar to those for
non-relativistic particles in the presence of a complex barrier, and
for convenience we use the language of wave mechanics to describe
them.  In particular, to the extent that we can ignore the imaginary
terms in the potential, these correspond to particles above a barrier
(since $\Lambda > 0$).  There are two caveats regarding discussing
this as a barrier penetration problem.  First, that our boundary
conditions do not correspond to the usual picture of incident,
reflected, and transmitted waves; in general, there are waves moving
in each direction on each side of the barrier.  Second, the pure
imaginary terms in the potentials play an extremely important role
here, even in the asymptotic regions.  These terms are needed not only
to represent nonadiabatic transitions, but also to keep up with the
local matter angle.

In this problem, then, the quantity of interest is not a reflection or
transmission coefficient, but rather $P(\nu_e \rightarrow \nu_e) =
|\Psi_e(x \rightarrow \infty)|^2$, the probability of the neutrino
being of the electron type far from the source.  In general, this is a
function of both source and detector positions, though typically only
the fully averaged result is presented.  However, those interference
terms could be important, and we present approximate expressions for
them in the next section.

Two well-known semiclassical treatments of this problem are via the
Wentzel-Kramers-Brillouin (WKB) \cite{Susyform2} and linear
Landau-Zener \cite{LZS,Haxton-LZ,Parke} methods.  The WKB technique
{\it globally} maps the ``potential'' discussed above onto the
free-particle potential (i.e., a constant density).  By ``global
mapping'', we mean a variable stretching of the axis that deforms the
shape of one potential into another.  In fact, the WKB treatment turns
out to be identical to the adiabatic approximation \cite{Susyform2}.
The linear Landau-Zener technique {\it locally} maps onto a linear
density (i.e., extends a linear profile from a single -MSW resonance-
point with the right density and derivative), and hence a
``potential'' with a parabolic real part and constant imaginary part.
While the linear Landau-Zener result is easy to derive and apply,
there are two problems.  First, it is notoriously difficult to get the
boundary conditions right (for a complete explanation of how to handle
this, see Ref.~\cite{Haxton-ELZ}).  Second, since Landau-Zener is a
point mapping, the expression for $P_{hop}$ is not very accurate.  In
the case of neutrino oscillations in of the sun, the exponential
Landau-Zener approximation circumvents these problems
\cite{Haxton-ELZ}.

The aim of this paper is to calculate the nonadiabatic corrections
semiclassically, but with a {\it global} mapping of the ``potential'',
where as in the Landau-Zener calculation we choose as a model the case
of a linear density.  By using a global mapping, the correct boundary
conditions are automatic. Further, the expression for
$P_{hop}$ is more accurate.  The approximate wave function is
uniformly valid in $x$ (though the approximation is not uniform in the
mixing parameters).


\section{Uniform Semiclassical Solution of the MSW Equations}


\subsection{Semiclassical Background}

In the adiabatic limit, only the lowest order is kept in the limit
$\lambda \rightarrow 0$ in Eq.~(\ref{eq:Hadi}), so the Hamiltonian is
taken as diagonal (no hopping from one mass eigenstate to the other)
and the integration is trivial.  The treatment at that order suggests
that in order to take into account the probability of hopping, we will
need to consider further orders in $\lambda$.  In this section, we
will show that it is possible to obtain a rather accurate expression
for the electron neutrino survival probability by making a
semiclassical expansion, i.e., by considering only the two lowest
orders in $\lambda$ when solving the MSW equations.

The expressions derived below will hold for values of the mixing
parameters from the extreme adiabatic limit up until the extreme
nonadiabatic limit.  In order to obtain solutions that hold in the
extreme nonadiabatic limit, one would formally have to consider all
orders in $\lambda$.  Since semiclassical expansions are asymptotic
(i.e., non-convergent) in general, it is not clear that this would
work in practice.  A much better approach for the extreme nonadiabatic
limit is to consider expansions in $1/\lambda$ \cite{Olivo}.

Semiclassical methods (for reviews, see Ref.~\cite{Semiclassical}) are
used in quantum mechanics to provide approximate solutions to the
Schr\"odinger equation in the limit that $\lambda$ is small.  As
noted, in the WKB method, one bases the approximate solutions on
free-particle solutions.  A procedure was developed by Miller and Good
\cite{Miller-Good} that instead bases the approximate solutions on the
known solutions of a Schr\"odinger equation with a similar potential.
In this method, the turning point singularities of the primitive WKB
method are regulated, and the solutions are uniformly valid: they hold
over the whole range in $x$ and are well-behaved at the turning
points.  A further advantage of the Miller-Good method is that
different potentials are treated with the same formalism, i.e., the
method of connection is the same for all potentials with the same
number and type of turning points.

The MSW equations of Eqs.~(\ref{eq:SE_e}) and (\ref{eq:SE_m}) are
Schr\"odinger-type equations for particles in the presence of complex
potentials of the form $V(x) = -\left[\varphi^2(x) \pm
i\lambda\eta\varphi'(x)\right]$, with $\varphi(x)$ independent of
$\lambda$.  In the Appendix, we summarize the extension of the uniform
semiclassical approximation to treat potentials with this specific
dependence on $\lambda$, originally introduced in
Ref.~\cite{Susyform1,Susyform2,Susyform3}.  This special form of the
potential arises in supersymmetric quantum mechanics; see
Ref.~\cite{Susyform3} for discussion.


\subsection{Application to the MSW Problem}

The method presented in the Appendix allows a uniform semiclassical
solution for $\Psi_e(x)$.  By ``uniform'', we mean that the local
error incurred by using the approximate solution developed there in
the exact differential equation is bounded as a function of $x$.  This
is to be distinguished from a semiclassical solution via the primitive
WKB method.  Such as solution has an unbounded error near a turning
point.  For MSW propagation, the turning points are complex (they are
near the resonance point).  In the nonadiabatic limit, the turning
points approach the real axis, which means that the WKB solutions are
unable to represent any of the nonadiabatic behavior.  In contrast,
the uniform semiclassical approximation used here is excellent for all
but the extreme nonadiabatic limit.  Since we will make a
semiclassical expansion, we explicitly show all factors of $\hbar$
(via $\lambda$).  Either an increasing or decreasing density can be
considered, by proper choice of $\eta$.

{}From the derivation in the Appendix, the general solution to
Eq.~(\ref{eq:SE_e}) is
\begin{equation}
\Psi_e(x) = K(x) \left[A D_{\nu}(z(x)) + B D_{\nu}(-z(x))\right]\,,
\end{equation}
where
\begin{equation}
\nu = \frac{\eta - 1 - i\Omega/\lambda}{2}\,,
\end{equation}
with
\begin{equation}
\Omega =
\frac{2i}{\pi} 
\int^{x_0^*}_{x_0}{dx \sqrt{\Lambda + \varphi^2(x)}}\,.
\end{equation}
The limits of integration above are the zeroes of the integrand,
chosen as described in the Appendix.  The argument of the parabolic
cylinder functions is given by
\begin{equation}
z(x) = \frac{1 + i}{\sqrt\lambda}\, S(x) \approx \frac{1 + i}{\sqrt\lambda}\,
(S_0(x) + \lambda S_1(x))\,,
\end{equation}
where $S_0(x)$ and $S_1(x)$ are described in the Appendix.

Given appropriate initial conditions, one can solve for $A$ and $B$.
In some situations, it may be useful to evaluate $\Psi_e(x)$ for all
$x$.  This requires evaluating the gamma function for complex argument
and the parabolic cylinder functions for general complex order and
argument. For general comments on routines available for the numerical
evaluation of special functions, see Ref.~(\cite{Lozier}).  The gamma
function for complex argument can be evaluated with {\sf CERNLIB}
\cite{Gamma}. General properties of the parabolic cylinder functions
may be found in Ref.~\cite{AS,WW,Dnu-gen,Dnu-pser}. While library
routines do exist for various special cases of the parabolic cylinder
functions, to our knowledge there is nothing available that is general
enough \cite{Dnu-num}.  The technique\footnote{The code is available
upon request from the authors.} used here is to use the power series
\cite{Dnu-pser} for small $|z|$, the asymptotic series \cite{WW} for
large $|z|$, and direct numerical integration of the defining
differential equation with {\sf ODEPACK} \cite{ODEPACK} for moderate
$|z|$.  Fortunately, one does not generally have to perform any
integrations for the parabolic cylinder functions, as only the
asymptotic forms are needed.

We will use the asymptotic forms at both the production and detection
points.  As shown below, this means that we assume adiabatic
propagation at those two points.  This matching is justified to the
extent that the production and detection points are sufficiently far
from the resonance point.  In practice, these requirements do not
present any difficulties.  Consider the sun as an example, with
neutrinos produced at the solar center.  If the resonance is near the
production point, or there is no resonance, then this implies that
$\delta m^2$ is large and the entire propagation is adiabatic, except
for extremely small mixing angles.  If the resonance is at a very low
density, i.e., approaching vacuum, then $\delta m^2$ is very small and
our approximation breaks down for other reasons described below.  Note
that in the linear Landau-Zener treatment, one has to handle the final
$x$ point carefully since the density runs negative at large $x$, and
$\theta(x) \rightarrow 0$, not $\theta_v$.  No such difficulties with
the boundary value of the matter angle arise in the treatment given
here.

The asymptotic forms developed below represent $\Psi_e(x)$ well for
large but finite $x$.  All of the expansions below are just to get
outside of the resonance region; we do not take $x$ so large that the
matter angle is either $\pi/2$ or $\theta_v$.  More precisely, (see
the discussion in the Appendix), the approximate solutions are
characterized by two scales, one set by $S_0(x)$ and the other by
$\varphi(x)$.  The function $S_0(x)$ is asymptotic outside the
resonance region, whereas $\varphi(x)$ is not asymptotic until the
density is zero or infinite.  In this formulation, $S_0(x)$, but not
$\varphi(x)$, will be taken to be asymptotic.  This means that we have
the control to connect opposite sides of the resonance region without
having to take $x \rightarrow \pm \infty$, i.e., we do not have to
extend the density profile indefinitely.


\subsection{Asymptotic Solutions and Connection Formulae}

Using the definition of the matter angle given in Eqs.~(\ref{eq:s2m})
and (\ref{eq:c2m}), we can rewrite the pre-exponential factors in the
asymptotic solution of $\Psi_e(x)$, Eq.~(\ref{eq:aPsi}).  There are
two cases, depending on how $\varphi$ and $\eta$ are chosen.  The
first case has $\varphi(x) = \cos{2\theta_v} - \zeta(x), \eta = -1$,
so
\begin{equation}
\left[\frac{\Lambda}{4\left(\Lambda + \varphi^2(x)\right)}\right]^{1/4}
\left(\frac{\varphi +
\sqrt{\Lambda + \varphi^2}}{\sqrt{\Lambda}}\right)^{+1/2}
= \cos\theta(x)\,,
\end{equation}
and
\begin{equation}
\left[\frac{\Lambda}{4\left(\Lambda + \varphi^2(x)\right)}\right]^{1/4}
\left(\frac{\varphi +
\sqrt{\Lambda + \varphi^2}}{\sqrt{\Lambda}}\right)^{-1/2}
= \sin\theta(x)\,.
\end{equation}
In the other case of $\varphi(x) = \cos{2\theta_v} - \zeta(x), \eta =
-1$, these are reversed.  In either case, the prefactors associated
with the various terms are:
\begin{eqnarray}
C_- & : & \cos\theta(x) \nonumber\\
D_- & : & \sin\theta(x) \nonumber\\
C_+ & : & \sin\theta(x) \nonumber\\
D_+ & : & \cos\theta(x) \,.
\end{eqnarray}
It is important to note that these will be evaluated at general values
of $x$ outside the resonance region; $|x|$ will not be so large that
$\theta(x) \rightarrow \pi/2$ or $\theta(x) \rightarrow \theta_v$.
The asymptotic wave functions are
\begin{equation}
\Psi_e(x \rightarrow -\infty)
= C_- \cos\theta(x) \exp\left(+\frac{i}{\lambda} I_p(x,x_i)\right)
+ D_- \sin\theta(x) \exp\left(-\frac{i}{\lambda} I_p(x,x_i)\right)
\end{equation}
and
\begin{equation}
\Psi_e(x \rightarrow +\infty)
= C_+ \sin\theta(x) \exp\left(-\frac{i}{\lambda} I_p(x,x_i)\right)
+ D_+ \cos\theta(x) \exp\left(+\frac{i}{\lambda} I_p(x,x_i)\right)\,.
\end{equation}
{}From their form, we see immediately that the asymptotic wave
functions represent adiabatic propagation.  The coefficients $C_\pm,
D_\pm$ still depend on $\eta$, which will allow us to consider
increasing or decreasing densities.  The phase integral function $I_p$
is defined in Eq.~(\ref{eq:Ip}).

With the asymptotic wave function in this form, it is rather easy to
apply the initial conditions.  As before, we take the neutrino at the
initial point $x_i$ to be a pure $\nu_e$, so
\begin{equation}
\Psi_e(x_i) = 1
\end{equation}
and
\begin{equation}
\Psi_e'(x_i) =
\left.\frac{\partial \Psi_e(x)}{\partial x}\right|_{x_i} =
-\frac{i}{\lambda} \eta\varphi(x_i)\,.
\end{equation}
In either case regarding the signs of $\varphi$ and $\eta$, one
immediately obtains $C_- = \cos\theta_i$ and $D_- = \sin\theta_i$, so
\begin{equation}
\Psi_e(x \rightarrow -\infty)
= \cos\theta(x) \cos\theta_i \exp\left(+\frac{i}{\lambda} I_p(x,x_i)\right)
+ \sin\theta(x) \sin\theta_i \exp\left(-\frac{i}{\lambda} I_p(x,x_i)\right)\,.
\end{equation}
which is of course the adiabatic solution given in Sec. (\ref{sec:adi}).

We now turn to the evaluation of the coefficients $C_+$ and $D_+$ that
are needed after the resonance.  {}From the above and
Eqs.~(\ref{eq:C-}) and (\ref{eq:D-}),
\begin{eqnarray}
C_- & = & \cos\theta_i =
C \exp\left(+\frac{i}{\lambda}{\rm Re} I_p(x_i,x_0)\right)
\left(A e^{-i\nu\pi} + B\right)\\
D_- & = & \sin\theta_i =
C \exp\left(-\frac{i}{\lambda}{\rm Re} I_p(x_i,x_0)\right)
\left(A e^{-i\nu\pi}\right)\,.
\end{eqnarray}
These determine the coefficients $A$ and $B$ of the general solution:
\begin{eqnarray}
A & = &
D^{-1} \exp\left(+\frac{i}{\lambda}{\rm Re} I_p(x_i,x_0)\right) e^{i\nu\pi}
\sin\theta_i \label{eq:A}\\
B & = &
C^{-1} \exp\left(-\frac{i}{\lambda}{\rm Re} I_p(x_i,x_0)\right)
\cos\theta_i -
D^{-1} \exp\left(+\frac{i}{\lambda}{\rm Re} I_p(x_i,x_0)\right)
\sin\theta_i \label{eq:B}\,.
\end{eqnarray}
Then
\begin{eqnarray}
C_+ & = & 2 i C D^{-1} e^{i\nu\pi} \sin\theta_i +
\exp\left(-\frac{2i}{\lambda}{\rm Re} I_p(x_i,x_0)\right) e^{-i\nu\pi}
\cos\theta_i\\
D_+ & = &  C^{-1} D e^{-i\nu\pi} \cos\theta_i -
\exp\left(+\frac{2i}{\lambda}{\rm Re} I_p(x_i,x_0)\right) e^{-i\nu\pi}
\sin\theta_i\,,
\end{eqnarray}
where $C$ and $D$ are given in Eqs.~(\ref{eq:C}) and (\ref{eq:D}).

The asymptotic forms of $\Psi_e(x)$ shown above are perfectly general,
and depend only on the assumption of adiabatic propagation outside the
resonance region.  The heart of this problem is the connection of the
asymptotic coefficients $C_-$ and $D_-$ to $C_+$ and $D_+$.  That
connection represents the integration of the solutions through the
resonance region.  In our case, that information is carried by the
coefficients $A$ and $B$ of the general (but approximate, due to the
mapping) solution in terms of parabolic cylinder functions.


\subsection{Resonance Transition Coefficients}

Above, the asymptotic wave functions were written in terms of the
adiabatic solutions, which is convenient for applying initial
conditions and deducing the connection formulae.  Before squaring the
asymptotic wave function to obtain the neutrino survival probability,
it is convenient to rewrite the wave function in a slightly different
form:
\begin{eqnarray}
\Psi_e(x \rightarrow +\infty) & = &
C_+ \sin\theta(x) \exp\left(-\frac{i}{\lambda} I_p(x,x_i)\right) +
D_+ \cos\theta(x) \exp\left(+\frac{i}{\lambda} I_p(x,x_i)\right)\nonumber\\
& = &
\left[
c_1 \sin\theta_i \exp\left(+\frac{i}{\lambda}{\rm Re} I_p(x_i,x_0)\right) +
c_2 \cos\theta_i \exp\left(-\frac{i}{\lambda}{\rm Re} I_p(x_i,x_0)\right)
\right]\nonumber\\
& & \times
\sin\theta(x) \exp\left(-\frac{i}{\lambda}{\rm Re} I_p(x,x_0)\right)\nonumber\\
& + &
\left[
d_1 \cos\theta_i \exp\left(-\frac{i}{\lambda}{\rm Re} I_p(x_i,x_0)\right) +
d_2 \sin\theta_i \exp\left(+\frac{i}{\lambda}{\rm Re} I_p(x_i,x_0)\right)
\right]\nonumber\\
& & \times
\cos\theta(x) \exp\left(+\frac{i}{\lambda}{\rm Re} I_p(x,x_0)\right)\,.
\end{eqnarray}
The terms inside the square brackets depend only on the source
position $x_i$, whereas the terms outside depend only on the final
position $x$.  Since all of the adiabatic phases and matter angles for
the asymptotic solutions are written explicitly, the matrix of
coefficients given by $c_1, c_2, d_1, d_2$ represents only the
nonadiabatic transitions in the resonance region.  These coefficients
change in the resonance region, but tend to asymptotically to
constants outside of it.  Since the $2 \times 2$ Hamiltonian is
Hermitian and traceless, the time-evolution operator must be a
member of the $SU(2)$ group.  Thus this matrix must assume the form
\begin{equation}
\left[\begin{array}{ll}
\phantom{-}c_1 & \phantom{-}c_2 \\
-c_2^* & \phantom{-}c_1^*
\end{array}\right]\,,
\end{equation}
where $|c_1|^2 + |c_2|^2 = 1$.

By comparison to the forms of $C_+$ and $D_+$ in Eqs.~(\ref{eq:C+}) and
(\ref{eq:D+}), the new coefficients are easily found to be
\begin{eqnarray}
c_1 & = & -\frac{\Gamma(-\nu)}{\sqrt{2\pi}}
\left(\frac{\Omega}{\lambda}\right)^{-i\Omega/2\lambda + \eta/2}
\left(\frac{e^{-i\pi/2}}{2}\right)^\nu
\frac{e^{-3i\pi/4}}{\sqrt{2}} \exp\left(+\frac{i\Omega}{2\lambda}\right)
2i\sin\left(\nu\pi\right)
\label{eq:c1}
\\
\nonumber\\
c_2 & = & e^{-i\nu\pi}
\label{eq:c2}
\\
\nonumber\\
d_1 & = & -\frac{\sqrt{2\pi}}{\Gamma(-\nu)}
\left(\frac{\Omega}{\lambda}\right)^{+i\Omega/2\lambda - \eta/2}
\left(\frac{e^{-i\pi/2}}{2}\right)^{-\nu}
e^{+3i\pi/4} \sqrt{2} \exp\left(-\frac{i\Omega}{2\lambda}\right)
e^{-i\nu\pi}\\
\nonumber\\
d_2 & = & -e^{-i\nu\pi}\,.
\end{eqnarray}
By analysis of two cases of $\eta = \pm 1$ separately, one can show
\cite{AS}:
\begin{equation}
|\Gamma(-\nu)|^2 \left(\frac{\Omega}{2\lambda}\right)^\eta =
\frac{\pi}{\sinh\left(\Omega\pi/2\lambda\right)}\,,
\end{equation}
\begin{equation}
\frac{2}{\pi} \, |\Gamma(-\nu)|^2 \left(\frac{\Omega}{2\lambda}\right)^\eta
|\sin(\nu\pi)|^2 e^{-\Omega\pi/2\lambda} =
1 - e^{-\Omega\pi/2\lambda}\,,
\end{equation}
where $\nu$ is given by Eq.~(\ref{eq:nu}).  Using these relations, it
may easily be verified explicitly that $d_1^* = c_1$ and $d_2^* = -
c_2$, and that
\begin{equation}
|c_1|^2 = 1 - e^{-\Omega\pi/\lambda}\,,
\end{equation}
\begin{equation}
|c_2|^2 = e^{-\Omega\pi/\lambda}\,,
\end{equation}
so $|c_1|^2 + |c_2|^2 = 1$.

Starting with Eq.~(\ref{eq:Hadi}), one can determine how $c_1$ and
$c_2$ depend on $\eta$, i.e., on whether the density is increasing or
decreasing.  One can show that $c_1$ must be independent of $\eta$,
and that $c_2$ must change sign if $\eta$ does.  With the present form
of $c_1$, this is not obvious.  Define the phase of $c_1$ as
\begin{equation}
c_1 = |c_1| e^{i\alpha}\,.
\end{equation}
An asymptotic series can be developed for this phase $\alpha$.  Using
the special form of the Stirling expansion of the gamma function for
purely imaginary argument given in Eqs.~(6.1.43-44) of Ref.~\cite{AS},
one can show that the phase of $c_1$, in the limit $\Omega/\lambda$ is
large, is
\begin{equation}
\alpha =
-\sum_{n=1}^{\infty} \frac{(-1)^{n-1} B_{2n}}{2n(2n-1)}
\left(\frac{2\lambda}{\Omega}\right)^{2n-1}\,,
\end{equation}
where $B_{2n}$ are Bernoulli numbers \cite{AS}.  When a linear density
is considered, this expression for the phase is equal to that given in
Ref.~\cite{Haxton-J}.  We do not use this limit for the phase in
general, since it requires that $\Omega/\lambda \gtrsim 1$, which is
unnecessarily restrictive on the range of validity of our main
approximation.  Since this is independent of $\eta$, so is $c_1$.
Note from the definition of $\nu$ in Eq.~(\ref{eq:nu}) that $c_2 =
e^{-i\nu\pi}$ is in fact a real number, though it may be positive or
negative, and changes sign if $\eta$ does.


\subsection{Calculation of $P(\nu_e \rightarrow \nu_e)$}

The electron neutrino survival probability at a general point $x$
after the resonance for a neutrino produced at $x_i$ is given by the
modulus squared of the amplitude $\Psi_e(x \rightarrow +\infty)$ for
the neutrino to be of the electron type.  First write
\begin{eqnarray}
\Psi_e(x \rightarrow +\infty) & = &
\left[
c_1 \sin\theta_i \exp\left(+\frac{i}{\lambda}{\rm Re} I_p(x_i,x_0)\right) +
c_2 \cos\theta_i \exp\left(-\frac{i}{\lambda}{\rm Re} I_p(x_i,x_0)\right)
\right]\nonumber\\
& & \times
\sin\theta(x) \exp\left(-\frac{i}{\lambda}{\rm Re} I_p(x,x_0)\right)\nonumber\\
& + &
\left[
c_1^* \cos\theta_i \exp\left(-\frac{i}{\lambda}{\rm Re} I_p(x_i,x_0)\right) -
c_2^* \sin\theta_i \exp\left(+\frac{i}{\lambda}{\rm Re} I_p(x_i,x_0)\right)
\right]\nonumber\\
& & \times
\cos\theta(x) \exp\left(+\frac{i}{\lambda}{\rm Re} I_p(x,x_0)\right)\,.
\end{eqnarray}

After taking the squared modulus of this expression for $\Psi_e(x)$,
and then reducing it, the survival probability takes the form
\begin{eqnarray}
P(\nu_e \rightarrow \nu_e)(x,x_i) & = &
\frac{1}{2} \left[1 + (1 - 2|c_2|^2)\cos{2\theta_i}\cos{2\theta(x)}\right]
\nonumber\\
& - & \frac{1}{2}|c_1|c_2 \sin{2\theta_i}\cos{2\theta(x)}
\cos\left(\frac{2}{\lambda}{\rm Re} I_p(x_i,x_0) + \alpha\right)
\nonumber\\
& + & \frac{1}{2} \sin{2\theta_i}\sin{2\theta(x)}
\cos\left(\frac{2}{\lambda}{\rm Re} I_p(x,x_i) - 2\alpha\right)
\nonumber\\
& - & |c_2|^2 \sin{2\theta_i}\sin{2\theta(x)}
\cos\left(\frac{2}{\lambda}{\rm Re} I_p(x_i,x_0) + \alpha\right)
\cos\left(\frac{2}{\lambda}{\rm Re} I_p(x,x_0) - \alpha\right)
\nonumber\\
& + & |c_1|c_2 \cos{2\theta_i}\sin{2\theta(x)}
\cos\left(\frac{2}{\lambda}{\rm Re} I_p(x,x_0) - \alpha\right)\,.
\end{eqnarray}
This, along with Eqs.~(\ref{eq:c1}) and (\ref{eq:c2}), is our main
result.  Recall that in our approximation, $c_2$ is real.  In general,
the phase $\alpha$ should be extracted from $c_1$ directly, rather
than taken from the asymptotic series for $\alpha$ given above.  This
simple form for the survival probability can be evaluated easily and
rapidly, providing accurate results for both the direct and
interference terms for all mixing parameters except for the extreme
nonadiabatic limit.  It is much more convenient than direct numerical
solution of the MSW equations, especially if many values of the mixing
parameters need to be explored.

When $\Omega/\lambda \gg 1$, i.e., the adiabatic limit, $|c_1|
\rightarrow 1$, $|c_2| \rightarrow 0$, $\alpha \rightarrow 0$, and
this general form for the survival probability reduces to
\begin{equation}
P(\nu_e \rightarrow \nu_e)(x,x_i) \rightarrow
\frac{1}{2} \left[1 + \cos{2\theta_i}\cos{2\theta(x)}\right]
+ \frac{1}{2} \sin{2\theta_i}\sin{2\theta(x)}
\cos\left(\frac{2}{\lambda}{\rm Re} I_p(x,x_i)\right)\,,
\end{equation}
which is the usual adiabatic result.

Typically, the final point $x$ will be taken in vacuum, so $\theta(x)
\rightarrow \theta_v$ and $\sqrt{\Lambda + \varphi^2(x)} \rightarrow
1$, and
\begin{equation}
\exp\left(\pm \frac{2 i}{\lambda}{\rm Re} I_p(x,x_0)\right)
= const. \times \exp\left(\pm 2 i x/\lambda\right)\,.
\end{equation}
(The same applies when the lower limit in $I_p$ is $x_i$, though the
constant will be different.)  The oscillation length in vacuum is $\pi
L$, where $L$ is given by Eq.~(\ref{eq:L}).  For example, in the solar
neutrino problem, the favored MSW parameters lead to an oscillation
length of $\approx 1000$ km \cite{Penn}.  In such cases, where the
oscillation length in vacuum is much less than the variation in the
source-detector distance, it will be appropriate to average over the
detector position.  If that is done, then the survival probability no
longer depends on $x$, but does still depend on $x_i$ and is given by
\begin{eqnarray}
P(\nu_e \rightarrow \nu_e)(x_i) & = &
\frac{1}{2} \left[1 + (1 - 2|c_2|^2)\cos{2\theta_i}\cos{2\theta_v}\right]
\nonumber\\
& - & \frac{1}{2}|c_1|c_2 \sin{2\theta_i}\cos{2\theta_v}
\cos\left(\frac{2}{\lambda}{\rm Re} I_p(x_i,x_0) + \alpha\right)\,.
\end{eqnarray}
This shows that the source term may be important even after detector
averaging.  If the source is extended, or if an energy spectrum is
considered, one can also average over the source position.  The
completely averaged result for the electron survival probability is
then given by
\begin{equation}
P(\nu_e \rightarrow \nu_e) =
\frac{1}{2} \left[1 + (1 - 2|c_2|^2)
\langle\cos{2\theta_i}\rangle_{src} \cos{2\theta(x)}\right]\,,
\label{eq:Pnu_da}
\end{equation}
where $\langle\cos{2\theta_i}\rangle_{src}$ indicates the average of
$\cos{2\theta_i}$ over the source position and energy spectrum.  In
the usual derivations, the source term is assumed to be averaged away,
yet no average over $\cos{2\theta_i}$ is shown.  However, one can get
away with this in some situations that suppress the source term
without any averaging over position or energy.

This structure for the fully-averaged survival probability is
completely general, and thus we interpret $|c_2|^2$ as the probability
of hopping from one mass eigenstate to the other in the passage
through the resonance region.  Thus
\begin{eqnarray}
P_{hop}
& = & |c_2|^2 = \exp\left(-\pi\Omega\right) \label{eq:Phop}\\
& = &
\exp\left(-2 i
\int^{x_0^*}_{x_0}{dx \sqrt{\zeta^2(x) - 2\cos{2\theta_v}\zeta(x) + 1}}\right)
\nonumber\\
& = &
\exp\left(-i \frac{\delta m^2}{2 E}
\int^{t_0^*}_{t_0}
{dt \left[\left[\frac{2 \sqrt{2} G_F E N_e(t)}{\delta m^2}\right]^2
- 2\cos{2\theta_v}\left[\frac{2\sqrt{2} G_F E N_e(t)}{\delta m^2}\right] 
+ 1\right]^{1/2}}\right)\nonumber\,.
\end{eqnarray}
This probability characterizes the non-adiabatic nature of the
evolution near the avoided level crossing; for purely adiabatic
evolution, $P_{hop} = 0$.  The limits of the integral are the complex
turning points of Eq.~(\ref{eq:SE_Psi}), i.e., the zeros of the
integrand, and are labeled such that ${\rm Im}\,x_0 > 0$.  The middle
form is particularly convenient since then the turning points are
located by $\zeta = \exp\left(\pm 2i\theta_v\right)$.  This result for
$P_{hop}$ is valid for both arbitrary mixing parameters and an
arbitrary monotonic density profile.  Since our solutions were based
on the solution for a linear density, the form of $P_{hop}$ follows
that for a linear density: a single exponential which vanishes in the
adiabatic limit.


\subsection{Comparisons of Different Densities}

For a linear density, we must recover the linear Landau-Zener result
\cite{Haxton-LZ,Parke}.  Equation (\ref{eq:Phop}) yields
\begin{equation}
P^{lin}_{hop} = \exp\left(-\pi\Omega^{lin}\right)\,,
\end{equation}
where
\begin{equation}
\Omega^{lin} = \frac{\gamma_c}{2} =
\frac{\delta m^2}{4 E} \frac{\sin^2{2\theta_v}}{\cos{2\theta_v}}
\left| \frac{1}{N(t)}\frac{d N(t)}{dt}\right|^{-1}_{res}\,,
\end{equation}
as expected.

For an exponential density, Eq.~(\ref{eq:Phop}) yields
\begin{equation}
P^{exp}_{hop} = \exp\left(-\pi\Omega^{exp}\right)\,,
\label{eq:Phop-exp}
\end{equation}
where
\begin{equation}
\Omega^{exp}
= \delta (1 - \cos{2\theta_v})
\end{equation}
and
\begin{equation}
\delta = \frac{\delta m^2}{2 E}
\left|\frac{1}{N(t)}\frac{dN(t)}{dt}\right|^{-1}\,.
\end{equation}
This is the leading exponential to the exact result for an exponential
density \cite{Haxton-ELZ,TP}.  The exact result is
\begin{equation}
P^{exp}_{hop} =
\frac{\exp\left(-\pi\delta (1 - \cos{2\theta_v})\right)
- \exp\left(-2\pi\delta\right)}
{1 - \exp\left(-2\pi\delta\right)}\,.
\end{equation}
Equation (\ref{eq:Phop-exp})for the exponential density was previously
obtained \cite{Pizzochero} by connecting the coefficients of the
coupled equations in the adiabatic basis through the complex plane
\cite{Landau}.  In Fig.~(1), we compare our fully-averaged result for
the survival probability in a exponential density (the parameters are
chosen to approximate the solar density \cite{Bahcall-dens}) with the
exact result.  The values of the vacuum angle chosen approximate those
of the two best-fit models for the MSW solution to the solar neutrino
problem \cite{Penn}.  In Figs.~(2) and (3), we show the accuracy of
our approximation by comparing our source term to the exact results.
The source term is defined as the survival probability, averaged over
detector, minus the survival probability, averaged over both source
and detector.  Note that in Eq.~(\ref{eq:Pnu_da}), when $\delta m^2/E$
is large, $\Omega$ is large, and $c_2 \rightarrow 0$, suppressing the
source term.  On the other hand, if the initial density is large
enough, then when $\delta m^2/E$ is small, $\theta_i \rightarrow
\pi/2$, and $\sin{2\theta_i} \rightarrow 0$, which also suppresses the
source term.  Therefore, the source term is non-zero only for
intermediate values of $\delta m^2$, as illustrated in Figs.~(2) and
(3).


\subsection{Breakdown of the Mapping}

As can be seen from Fig.~(1), our approximation does not hold in the
extreme nonadiabatic limit, where $\delta m^2 \rightarrow 0$.  As
emphasized in Ref.~\cite{Pechukas}, the Miller-Good method only works
well when the mapping is invertible.  Given two potentials $V_A$ and
$V_B$, the mapping is good only if it makes as much sense to map $V_A
\rightarrow V_B$ as $V_B \rightarrow V_A$.  If this is not true, then
the mapping is a projection, and something is lost.  Invertibility may
be thus be associated with a ``sameness of topology.''  More
precisely, when the mapping is not invertible, the comparison
potential becomes multivalued.

Let us consider the treatment of an exponential density.  In this
case, the root of the failure in mapping is the difference in the
topology of higher-order turning points of the two potentials
corresponding to linear and exponential densities.  The turning points
are located by $\zeta = \exp\left(\pm 2i\theta_v\right)$.  For a
linear density, there are only two turning points.  For an exponential
density, however, additional, higher-order turning points can be found
by the transformation $x \rightarrow x + 2\pi n x_s$, where $n$ is an
integer and $x_s$ is the scale height of the exponential in our
dimensionless units.  As noted in the Appendix, we considered only the
primary turning points, i.e., those closest to the real axis.  When
only the lowest-order turning points of the exponential density are
taken into account, then the two potentials can be made only
approximately equivalent.  In principle, the way to cure this problem
is to use an comparison potential with the same number (infinite, if
necessary) of turning points as the original one.  In practice, this
may be rather cumbersome.

Consider how a path in the $x$-plane is mapped into the $S$-plane.  In
the Appendix, we discuss why the locations of the primary turning
points are only considered to lowest order in $\lambda$.  In
particular, the turning points in the $S$-plane are located by $S(x_0)
\approx S_0(x_0) = +i\sqrt\Omega$, $S^*(x_0) \approx S_0^*(x_0) =
-i\sqrt\Omega$.  The resonance point $x_c$ is mapped to $S(x_c)
\approx 0$.  In the extreme adiabatic limit, the path from $-\infty$
to $+\infty$ along the real axis in the $x$-plane is mapped onto a
path from $-\infty$ to $+\infty$ along the real axis in the $S$-plane.
As the mixing parameters become more and more nonadiabatic, the path
in the $S$-plane makes more and more of an excursion into one half
(upper or lower) complex plane near the resonance.  At $\pm\infty$, it
returns to the real axis.  In both planes, the paths run between the
primary turning points.  In the extreme nonadiabatic limit, however,
the path of $S(x)$ eventually crosses a turning point.  There is
then a topological difference between the two planes -- in one case,
the path runs between the primary turning points, and in the other, it
does not.  Because of how the turning points are anchored, this
indicates that the mapping has folded the complex plane over, and the
comparison potential is multivalued.

The need to impose the same turning point topology between the
original and comparison potentials restricts the applicability of
Eq.~(\ref{eq:Phop}) to monotonically-varying electron densities, i.e.,
those with a single MSW resonance.  If there are two or more close MSW
resonances, one cannot use a linear density to construct the
comparison potential. Such situations are considered in
Ref. \cite{Haxton-J,Noise1}.


\section{Concluding Remarks}

We have studied a uniform semiclassical approximation for the
matter-enhanced neutrino oscillations for two flavors, assuming a
monotonically changing but otherwise arbitrary density profile.  We
obtained an analytic expression for the electron neutrino survival
probability, unaveraged over either detector or source positions.  Our
result is valid for a large range of the mixing parameters, up to but
not including the extreme nonadiabatic limit.  Upon averaging over
detector and source positions, we recover expressions previously
obtained in the literature.  Since our expressions are valid for
arbitrary densities, they may be applied not only to the sun, but to
all settings in which resonant neutrino conversion can occur, such as
supernovae and the early universe.

The method of analytic continuation utilized in Ref.~\cite{Pizzochero}
for an exponential density was extended in Ref.~\cite{Kuo}, where the
general form of Eq.~(\ref{eq:Phop}) for an arbitrary monotonic density
profile was found.  Results for several other analytically solvable
densities are presented there.  Our analysis not only yields an
expression for the hopping probability which coincides with
Ref.~\cite{Kuo}, but also provides the source and detector terms.

As noted, we assumed a monotonic density profile, so this formalism is
not suitable for studying neutrino propagation in stochastic media
(e.g., with density fluctuations), as has recently been studied for
the sun and type-II supernovae in Ref.~\cite{Noise1,Noise2}.


\section*{ACKNOWLEDGMENTS}

We thank J.M. Fetter and R.E. Meyer for helpful discussions.  This
research was supported in part by the U.S. National Science Foundation
Grant No.\ PHY-9314131 at the University of Wisconsin, and in part by
the University of Wisconsin Research Committee with funds granted by
the Wisconsin Alumni Research Foundation.


\newpage
\appendix
\section*{Supersymmetry-Inspired Uniform Approximation}


\subsection{General Treatment}

Consider the Schr\"odinger-type equation
\begin{equation}
-\lambda^2\,\frac{\partial^2 \Psi(x)}{\partial x^2} -
\left[\Lambda + \varphi^2(x) + i\lambda\eta\varphi'(x)\right]\Psi(x) = 0 \,,
\label{eq:SE_Psi}
\end{equation}
where $\Lambda$ is a real, positive constant, and $\varphi$ is a real,
monotonic function on the real axis, and is analytic in the complex
plane.  In the above and what follows, everything is dimensionless,
and $\lambda$ is being used as a placeholder for $\hbar$.  Neither
$\Lambda$ nor $\varphi$ depends on $\lambda$.  We will solve this
equation in an approximation that treats $\lambda$ as formally small.
In the physical problem represented by Eq.~(\ref{eq:SE_Psi}), the
variable $x$ is real.  However, for addressing the mathematical
question of the solution of this differential equation, we consider
$x$ to be complex.  We assume\footnote{The case in which the zeros are
on the real axis, while not relevant here, can be treated similarly to
the rest of the Appendix; see Ref.~\cite{Susyform1}.} that $\Lambda +
\varphi^2(x)$ has two zeros in the complex plane, i.e., points
$x_0,x_0^*$ where $\varphi= \pm i\sqrt{\Lambda}$.  These points are
taken to be the turning points of Eq.~(\ref{eq:SE_Psi}).  We will
discuss below why the turning points can be taken at lowest order,
i.e., given as the zeros of $\Lambda + \varphi^2(x)$, rather than of
$\Lambda + \varphi^2(x) + i\lambda\eta\varphi(x)$.  In general, there
can be more than two zeros of $\Lambda + \varphi^2$; for now, we only
consider the two closest to the real axis, and label them so that
${\rm Im}\,x_0 > 0$.  The overall sign on $\varphi$ is chosen to make
${\rm Im}\,\varphi(x_0) > 0$, and $\eta$ is taken to be $\pm 1$ as
needed so that $\varphi(x)$ has the desired sign.  If presented with
an equation like Eq.~(\ref{eq:SE_Psi}), but with the opposite sign on
the imaginary term, one can always conjugate it and solve as below for
$\Psi^*(x)$, so the treatment here is general.

We will map Eq.~(\ref{eq:SE_Psi}) onto the comparison equation
\begin{equation}
-\lambda^2\,\frac{\partial^2 U(S)}{\partial S^2} -
\left[\Omega + S^2 + i\lambda\eta\right]U(S) = 0\,,
\label{eq:SE_U}
\end{equation}
where $\Omega$ is a real, positive constant (and will be determined
below).  This equation, considered as a function of $S$, also has two
conjugate turning points in the complex plane.  By ``map'', we mean
that we will find a change of variables $S = S(x)$ such that the
potential in the comparison equation is deformed into the potential in
the original equation.  That statement indicates how the real axis
will be stretched.  However, we will also have to consider how the
complex $x$-plane is mapped onto the complex $S$-plane.  In
particular, the turning points in the $x$-plane must be mapped onto
the turning points in the $S$-plane.  The comparison equation is
chosen to be one for which exact analytic solutions are known, and
which is as similar as possible to the original equation.  If we
require $\varphi(x)$ to be monotonic for real $x$, then imaginary term
in the comparison equation may be taken as constant.  Other than
$\Omega$, this comparison equation is taken with no free parameters;
such parameters can always be scaled away, and so are irrelevant here.
That the turning-point topologies of the original and comparison
problems be the same is critical to the method.  In this case, we are
mapping an as-yet unspecified real barrier onto a parabolic barrier,
and the imaginary term onto an imaginary constant.  However, we can
map onto any convenient potential with known solutions.

In principle, if we could find the change of variables $S = S(x)$
exactly, then we would have exact solutions for $\Psi(x)$ in terms of
the known functions $U(S(x))$.  In general, the solution for the
change of variables $S = S(x)$ would be at least as difficult as
direct solution of the original problem.  The approximation made to
solve Eq.~(\ref{eq:SE_Psi}) will be to approximate $S(x)$ as a
truncated power series in $\lambda$.  This method of uniform
approximation via mapping is also known as the method of comparison
equations (note Ref.\cite{Langer}).  The work here was inspired by the
ideas of Miller and Good \cite{Miller-Good}.  For further work on the
theory of their method, see also
Ref.~\cite{Miller-Good:more,Pechukas}, and the related works in
Ref.~\cite{Miller-Good:similar}.

In the original Miller-Good problem, the imaginary terms in the
potentials above are not present.  They treat the cases of a particle
bound in a well and traveling in the presence of a barrier, mapping
onto a parabolic well and barrier, respectively, each of which have as
solutions the parabolic cylinder functions.  In their formalism, one
immediately sees that the WKB approximation amounts to mapping onto
the free-particle potential; the mismatch in turning-point topologies
is the origin of the failure of the WKB method (via the zeros in what
is essentially a Jacobian, see Eq.~(\ref{eq:K}) near the turning
points.  With the Miller-Good formalism, the wave function is
continuous through the turning points.

The notation used here has some important differences from this
previous work.  This allows some difficulties and errors to be
resolved.  In particular, $\eta$ will be used differently here.  We
continue to take $\varphi(x)$ to be real on the real axis and to be
monotonic.  In those papers, it was assumed that $\varphi(x)$
monotonically increasing along the real axis would imply ${\rm
Im}\,\varphi(x_0) > 0$.  While this is suggested by the Cauchy-Riemann
conditions applied to $\varphi(x)$ on the real axis, it need not
always be true.  No such assumption is made here.  For each density,
one simply has to make sure that the signs of $\varphi(x)$ and $\eta$
are defined so that $\eta\varphi(x)$ represents the right physics and
that ${\rm Im}\,\varphi(x_0) > 0$.

This mapping will be accomplished as
\begin{equation}
\Psi(x) = K(x)U(S(x))\,,
\end{equation}
where the form of $K(x)$ will be chosen and $S(x)$ will be defined by
that choice.  Using this form of $\Psi(x)$ and Eq.~(\ref{eq:SE_U}), we
can rewrite Eq.~(\ref{eq:SE_Psi}).  By making the choice
\begin{equation}
K(x) = \frac{1}{\sqrt{\vphantom{\dot S_0}S'(x)}}\,,
\end{equation}
and dividing through by $\Psi$, we find
\begin{equation}
\lambda^2\,\frac{K''}{K} -
S'\,^2 \left[\Omega + S^2 + i\lambda\eta\right] +
\left[\Lambda + \varphi^2 + i\lambda\eta\varphi' \right] = 0\,.
\label{eq:DE_S}
\end{equation}
So far, no approximation has been made, and the form
\begin{equation}
\Psi(x) = \frac{1}{\sqrt{\vphantom{\dot S_0} S'(x)}} U(S(x))
\end{equation}
is a purely formal solution of Eq.~(\ref{eq:SE_Psi}) in terms of the
solutions of Eq.~(\ref{eq:SE_U}).  If we could find the change of
variables $S = S(x)$ exactly, then we would have exact solutions for
$\Psi(x)$ in terms of the known functions $U(S(x))$.  In general, the
solution for the change of variables $S = S(x)$ would be at least as
difficult as direct solution of the original problem.  To avoid that,
we will approximate $S(x)$.  Very crudely, this procedure is a
perturbation expansion in the shapes of the two potentials --- the
more they resemble each other, the more our approximation to the
change of variables is justified, and the better our solutions
$\Psi(x)$ will be.  Since $\varphi$ is independent of $\lambda$, all
of the $\lambda$ dependence in Eq.~(\ref{eq:DE_S}) is explicit.  We
expand $S(x)$ in powers of $\lambda$:
\begin{equation}
S(x) = S_0(x) + \lambda S_1(x) + \ldots \,.
\end{equation}
The power of this method is that we can obtain a good solution by
keeping only the semiclassical terms (the lowest two orders in
$\lambda$).  In the original Miller and Good problem
\cite{Miller-Good}, the $i\lambda\varphi'$ term was not present in the
potential.  Therefore, only $\lambda^2$ appears in
Eq.~(\ref{eq:DE_S}), and one can expand in $\lambda^2$ instead of
$\lambda$, which leads to $S(x) \approx S_0(x) + {\cal O}(\lambda^2)$,
making solving for the mapping quite simple.  In our case, since
$\lambda$ appears directly in Eq.~(\ref{eq:DE_S}), we must expand in
$\lambda$, which leads to
\begin{equation}
S(x) \approx S_0(x) + \lambda S_1(x)\,,
\end{equation}
which makes solution of the mapping somewhat more complicated, but
still much easier to solve than the original equation.  After
expansion of Eq.~(\ref{eq:DE_S}) in $\lambda$, we group by order in
$\lambda$ and demand that each order vanish independently, as
$\lambda$ is a free parameter as far as the mathematics are concerned.
This yields the equations:
\begin{equation}
\begin{array}{lrcl}
\hbox{${\cal O}(\lambda^0)$\,:}
& (\Lambda + \varphi^2) & = & S_0'^2 (\Omega + S_0^2) \\
\hbox{${\cal O}(\lambda^1)$\,:}
& i\eta\varphi' & = & 2(\Omega + S_0^2) S_0' S_1' +
S_0'^2 (i\eta + 2 S_0 S_1)\,.
\end{array}
\label{eq:DE_S0S1}
\end{equation}
While the original equation to be solved was linear, after
approximation the system of equations to be solved is nonlinear.  In
particular, the equation for the ${\cal O}(\lambda^2)$ terms, which
involves $S_2(x)$, is probably analytically intractable.
Nevertheless, the integrations for $S_0(x)$ and $S_1(x)$ can be
performed, and the results are given below.  In those integrations,
the branch cut for the logarithm and square-root functions is taken
along the negative real axis.  Before solving for $S_0(x)$ and
$S_1(x)$, we show what will be left over.  Using the relations for
$S_0(x)$ and $S_1(x)$, one can show
\begin{eqnarray}
-\lambda^2\,\frac{\partial^2 \Psi_{appr}(x)}{\partial x^2} -
\left[\Lambda + \varphi^2(x) + i\lambda\eta\varphi'(x) +
\lambda^2\epsilon(x)\right] \Psi_{appr}(x) = 0 \,.
\end{eqnarray}
In the rest of the Appendix, $\Psi(x)$ always denotes the approximate
wave function, and we drop the subscript.  The degree to which the
approximate solution fails to solve the exact differential equation is
the local error, and is of the form
\begin{eqnarray}
\epsilon(x) & = &
\frac{3}{4}\left(\frac{S_0''}{S_0'}\right)^2
- \frac{1}{2}\left(\frac{S_0'''}{S_0'}\right)
- S_0'^2 S_1^2 - 2 S_0' S_1' (i\eta + 2 S_0 S_1)
- S_1'^2 (\Omega + S_0^2) \nonumber\\
& + & (\mbox{terms that depend on } S_2)\,.
\end{eqnarray}
The first two terms are familiar from either the WKB \cite{WKB-error}
or Miller-Good \cite{Miller-Good} problems.\footnote{While these
methods have the same form for the local error, the global results can
be rather different, e.g., the transmission coefficient
\cite{Miller-Good}.  Note that the WKB error term is singular at the
turning points, whereas the Miller-Good error term is bounded.}  The
next three terms arise from the more general form of the potential
considered here. Unfortunately, the remaining terms depend on $S_2$,
for which we have no analytic solution.

The turning points of the original and comparison equations are taken
to be the zeros of $(\Lambda + \varphi^2(x))$ and $(\Omega + S^2(x))$,
respectively.  Since these are real for real $x$, the turning points
are complex conjugates.  As noted above, the turning points of
original equation are labeled so that ${\rm Im}\,x_0 > 0$, and the
sign of $\varphi(x)$ is chosen so that $\varphi(x_0) = i\sqrt\Lambda$.
We map the turning points of the original equation onto the turning
points of the comparison equation by demanding $S(x_0) =
i\sqrt\Omega$.  The way this correspondence is made ensures that the
mapping does not flip the complex plane about the real axis (it is not
flipped about the imaginary axis either, as will be shown below).
These choices make it easier to avoid integration errors below.  Note
that all of the turning points are treated only at lowest order.

The formal solution for $S_0(x)$ can be written immediately from
Eq.~(\ref{eq:DE_S0S1}), including the turning-point correspondence
condition of $S_0(x_0) = i\sqrt\Omega$:
\begin{equation}
I_p(x,x_0) \equiv
\int^x_{x_0} {dx \sqrt{\Lambda + \varphi^2(x)}} =
\int^{S_0(x)}_{i\sqrt\Omega} {dS_0 \sqrt{\Omega + S_0^2}}\,.
\label{eq:Ip}
\end{equation}
In order to evaluate $S_0(x)$, we will first need $\Omega$, the energy
of the comparison system.  This is determined by demanding that the
conjugate turning points correspond, i.e., $S_0(x_0^*) =
-i\sqrt\Omega$.  When both sides of Eq.~(\ref{eq:Ip}) are integrated
between their turning points, the right-hand side can be done
explicitly, yielding
\begin{equation}
-\frac{i\Omega\pi}{2} =
\int^{x_0^*}_{x_0}{dx \sqrt{\Lambda + \varphi^2(x)}}\,.
\label{eq:Omega}
\end{equation}
With $\Omega$ determined, an implicit solution for $S_0(x)$ can be
obtained through integration of Eq.~(\ref{eq:Ip}) to a general point
$x$:
\begin{equation}
I_p(x,x_0) =
-\frac{i\Omega\pi}{4} + \frac{S_0}{2}\sqrt{\Omega + S_0^2} +
\frac{\Omega}{2}
\ln\left(\frac{S_0 + \sqrt{\Omega + S_0^2}}{\sqrt\Omega}\right)\,.
\label{eq:S0}
\end{equation}
The fact that the solution for $S_0(x)$ is left in this implicit form
does not present any difficulties.  When the asymptotic forms are
used, this expression can be solved approximately.  If the full forms
of the parabolic cylinder functions are being used, then one will be
taking a numerical approach anyway, and the solution for $S_0(x)$ is
rather easy.  Using the Schwarz Reflection Principle and the integrals
for $S_0(x)$ and $\Omega$, one can show that $S_0(x)$ is real for real
$x$.  Then $I_p(x,x_0)$ separates into real and imaginary parts as
follows:
\begin{equation}
I_p(x,x_0) =
{\rm Re} (I_p(x,x_0)) - \frac{i\Omega\pi}{4}\,.
\label{eq:Ip_ri}
\end{equation}
This will be needed to show that the exponentials in the asymptotic
solution have purely imaginary arguments.

To solve for $S_1(x)$, we use the first of Eq.~(\ref{eq:DE_S0S1}) to
rewrite the second as
\begin{equation}
\frac{i\eta}{2}\frac{\varphi'}{\sqrt{\Lambda + \varphi^2}} =
S_1'\sqrt{\Omega + S_0^2} +
\frac{1}{2}\frac{S_0'}{\sqrt{\Omega + S_0^2}}
\left(i\eta + 2 S_0 S_1\right)\,.
\end{equation}
We integrate from $x_0$ to $x$, changing variables as needed
\begin{equation}
\frac{i\eta}{2}
\int^{\varphi(x)}_{\varphi(x_0)}
{\frac{d\varphi}{\sqrt{\Lambda + \varphi^2}}}
=
\int^x_{x_0}
{dx \frac{d}{dx}(S_1 \sqrt{\Omega + S_0^2})}
+
\frac{i\eta}{2}
\int^{S_0(x)}_{S_0(x_0)}
{\frac{dS_0}{\sqrt{\Omega + S_0^2}}}\,.
\end{equation}
In all three integrals, the lower limit has a positive imaginary part
and the upper limit is real; this ensures that all of the square root
functions are on the same sheet.  Upon substitution of the limits of
integration, this yields
\begin{equation}
S_1(x) = \frac{i\eta}{2\sqrt{\Omega + S_0^2}}
\left\{
\ln{\left[\frac{\varphi + \sqrt{\Lambda + \varphi^2}}{\sqrt{\Lambda}}\right]}
+
\ln{\left[\frac{\sqrt{\Omega}}{S_0 + \sqrt{\Omega + S_0^2}}\right]}
\right\}\,,
\label{eq:S1}
\end{equation}
which is purely imaginary for real $x$, as claimed in
Ref.~\cite{Susyform2,Susyform3}.  In those references, there were
typographical errors (corrected here), and this was not obvious.  In
order to get $S_1$ right (i.e., pure imaginary), one has to be careful
about a branch cut crossing during the integration.  In order to
facilitate that, $\varphi$ is defined here with the sign noted in
Eq.~(\ref{eq:varphi}).

The prefactor $K(x)$ is given by
\begin{equation}
K(x) = \frac{1}{\sqrt{\vphantom{\dot S_0}S'(x)}} \approx
\frac{1}{\sqrt{S_0'(x)}} =
\left[\frac{\Omega + S_0^2(x)}{\Lambda + \varphi^2(x)}\right]^{1/4}\,.
\label{eq:K}
\end{equation}
The truncation of $S'$ at lowest order here is consistent with our
overall policy of keeping the two lowest orders in $\lambda$, and will
be justified below.  This form allows us explain why the turning
points may be taken at lowest order.  Consider how the exact change of
variables $S = S(x)$ would work at a single point, where the local
momenta in both problems could be taken as constant.  Then the change
of variables is just a scale change, and $K^2(x)$ is the ratio of the
exact local momenta, the zeros of which are the exact turning points.
When we approximate $S(x)$ by power series in $\lambda$, $K^2(x)$ will
be the ratio of the local momenta, taken at the proper order, the
zeros of which are the turning points, taken at the proper order.
{}From the expression for $K(x)$ above, this indicates that the
turning points should be taken at the lowest order in $\lambda$.  The
denominator of $K(x)$ vanishes at the turning points.  For there to be
any hope of of $\Psi(x)$ being well-behaved at the turning points, the
numerator must vanish also.  This is why we demanded the turning-point
matching as above.  In the WKB case, the numerator never vanishes, and
the connection between the mismatch in turning point topologies and
the singularity of the WKB solutions at the turning points is evident.
When the turning points are matched properly, one can show that $K(x)$
is well-behaved for all $x$.  The method of proof is to expand
Eq.~(\ref{eq:Ip}) near one turning point; this reveals that $K(x)$
tends to a constant as the turning point is approached.  The same
holds for the other turning point as well.

After successively solving for $\Omega$, $S_0(x)$, $S_1(x)$, and
$K(x)$, one can write the approximate wave function as
\begin{equation}
\Psi(x) =
\left[\frac{\Omega + S_0^2(x)}{\Lambda + \varphi^2(x)}\right]^{1/4}
U(S_0(x) + \lambda S_1(x))\,,
\end{equation}
where $U(S_0(x) + \lambda S_1(x))$ approximately solves
Eq.~(\ref{eq:SE_U}), the comparison equation, and $\Psi(x)$
approximately solves Eq.~(\ref{eq:SE_Psi}), the original
Schr\"odinger-type equation.  By taking
\begin{equation}
z(x) = \frac{1 + i}{\sqrt\lambda}\, S(x) \approx \frac{1 + i}{\sqrt\lambda}\,
(S_0(x) + \lambda S_1(x))\,,
\label{eq:z}
\end{equation}
one can show that Eq.~(\ref{eq:SE_U}) is Weber's equation
\cite{AS,WW,Dnu-gen} for the parabolic cylinder function
$D_{\nu}(z)$, where the order $\nu$ is given by
\begin{equation}
\nu = \frac{\eta - 1 - i\Omega/\lambda}{2}\,.
\label{eq:nu}
\end{equation}
The general solution may be taken to be
\begin{equation}
\Psi(x) = K(x) \left[A D_{\nu}(z(x)) + B D_{\nu}(-z(x))\right]\,.
\label{eq:gPsi}
\end{equation}
There are four solutions to Weber's equation, any two of which can be
taken as independent; the two above are convenient.  Given appropriate
initial conditions, one can solve for $\Psi(x)$ everywhere.  The
coefficients $A$ and $B$ will be determined below by applying the
initial conditions to the asymptotic form of the solution.  The
evaluation of the parabolic cylinder functions is discussed in the
body of the paper.


\subsection{Asymptotic Forms}

In many cases, one only needs the wave function as $x \rightarrow \pm
\infty$ (the reasons for this in the MSW problem are explained in the
body of the paper).  In this section, we develop the asymptotic forms
of $\Psi(x)$, which are easier to work with than the general form
above.  Another benefit of the asymptotic forms is that it becomes
much easier to count powers of $\lambda$, thus ensuring that we are
working to a consistent order.

Using the definitions in the previous section, one can easily show
\begin{equation}
S_0(x) \mathop{\longrightarrow}\limits_{x \rightarrow \pm \infty} \pm \infty
\end{equation}
and
\begin{equation}
S_1(x) \mathop{\longrightarrow}\limits_{x \rightarrow \pm \infty} 0\,.
\end{equation}
With these crude limits, and the fact that $S_0(x)$ is real for real
$x$, we can determine the phases to the arguments of $D_{\nu}(z(x))$
and $D_{\nu}(-z(x))$ to be $\pi/4$ and $-3\pi/4$, respectively for $x
\rightarrow + \infty$, and vice versa for $x \rightarrow - \infty$.
We use $-3\pi/4$ instead of $5\pi/4$ to stay inside the principal
branches of the square root and logarithm functions used below.  In
particular, one must be careful when rewriting $z^\nu$.  The
asymptotic forms of the parabolic cylinder functions for $|z| >>
|\nu|$ are given in Ref.~\cite{WW}.  For $-3\pi/4 < {\rm arg}(z) <
3\pi/4$,
\begin{equation}
D_{\nu}(z)\,
\mathop{\longrightarrow}\limits_{z \rightarrow \infty}\,
\exp\left(-\frac{z^2}{4}\right) z^{\nu}
\left(1 + {\cal O}\left(\frac{1}{z^2}\right) \right)
\label{eq:aDnu1}
\end{equation}
and for $-5\pi/4 < {\rm arg}(z) < -\pi/4$,
\begin{eqnarray}
D_{\nu}(z)\,
& \mathop{\longrightarrow}\limits_{z \rightarrow \infty}\, &
\exp\left(-\frac{z^2}{4}\right) z^{\nu}
\left(1 + {\cal O}\left(\frac{1}{z^2}\right) \right) \nonumber \\
& - & \frac{\sqrt{2\pi}}{\Gamma(-\nu)}
e^{-i \nu \pi} \exp\left(\frac{z^2}{4}\right) z^{-\nu -1}
\left(1 + {\cal O}\left(\frac{1}{z^2}\right) \right)\,.
\label{eq:aDnu2}
\end{eqnarray}
In the common range of validity, the difference between the two forms
is negligibly small.

These asymptotic forms make the dependence upon $\lambda$ in the
various terms easy to see, and show how to keep consistent orders in
$\lambda$.  As befits our semiclassical expansion, we only keep the
lowest two orders in $\lambda$ in the exponentials of the asymptotic
wave functions.  Using the form of $z$ given in Eq.~(\ref{eq:z}),
\begin{equation}
\exp\left(\pm \frac{z^2}{4} \right) =
\exp\left(\pm \frac{i}{2} \left(\frac{S_0^2}{\lambda} + 2 S_0 S_1 +
{\cal O}(\lambda)\right)\right)\,,
\end{equation}
\begin{equation}
z^\nu =
\left(\frac{1 + i}{\sqrt\lambda}\right)^\nu S_0^\nu
\exp\left(-\frac{i\Omega}{2}\frac{S_1}{S_0} + {\cal O}(\lambda)\right)\,,
\end{equation}
\begin{equation}
z^{-\nu -1} =
\left(\frac{1 + i}{\sqrt\lambda}\right)^{-\nu -1} S_0^{-\nu -1}
\exp\left(+\frac{i\Omega}{2}\frac{S_1}{S_0} + {\cal O}(\lambda)\right)\,,
\end{equation}
\begin{equation}
1 + {\cal O}\left(\frac{1}{z^2}\right) =
\exp\left({\cal O}(\lambda)\right)\,,
\end{equation}
\begin{equation}
K(x) = \frac{1}{\sqrt{S'(x)}} =
\frac{1}{\sqrt{S_0'(x)}} \exp\left({\cal O}(\lambda)\right)\,.
\end{equation}

Now we expand the various pieces for large $|x|$.  Using
Eq.~(\ref{eq:S0}), and keeping only terms growing or constant in
$|S_0(x)|$, one can easily show
\begin{equation}
\frac{S_0^2(x)}{2\lambda}
\mathop{\longrightarrow}\limits_{x \rightarrow \pm \infty}
\pm \frac{I_p(x,x_0)}{\lambda} + \frac{(\pm i\pi - 1)\Omega}{4\lambda} +
\frac{\Omega}{4\lambda} \ln{\left(\frac{\Omega}{\lambda}\right)} -
\frac{\Omega}{2\lambda} \ln{\left(\frac{2|S_0(x)|}{\sqrt\lambda}\right)}\,.
\label{eq:aS0}
\end{equation}
When $|S_0| >> \Omega$ and this expansion is valid, then $|z| >>
|\nu|$ and the $D_{\nu}$'s can be put in their asymptotic forms.
Using Eq.~(\ref{eq:Ip_ri}), one can see that $S_0^2$ is purely real
for real $x$, as claimed.  The various $\lambda$ terms were introduced
to show where $\hbar$'s would appear if we took these equations out of
dimensionless form.  The expansion for $S_1$ from Eq.~(\ref{eq:S1}) is
straightforward, and yields
\begin{equation}
S_1(x)
\mathop{\longrightarrow}\limits_{x \rightarrow \pm \infty}
\frac{i\eta}{2|S_0|}
\left\{
\ln{\left[\frac{\varphi + \sqrt{\Lambda + \varphi^2}}
{\sqrt{\Lambda}}\right]}
\mp \ln{\left(\frac{\sqrt\lambda}{\sqrt\Omega}\right)}
\mp \ln{\left(\frac{2|S_0|}{\sqrt\lambda}\right)}
\right\}\,.
\label{eq:aS1}
\end{equation}
In this expansion, $|S_0|$ was treated asymptotically but $|\varphi|$
was not.  (At the MSW resonance point discussed in the body of this
paper, $\varphi(x_c) = 0$ but $S_0(x_c) \neq 0$ in general.)  The
nonadiabatic region is in general far narrower than the region over
which the matter angle is varying appreciably.  The region in which
$S_0$ cannot be treated asymptotically is essentially the nonadiabatic
region, whereas $\varphi$ cannot be treated asymptotically until the
matter angle is very close to either $\pi/2$ or $\theta_v$.  We expand
$K(x)$ as
\begin{equation}
K(x)
\mathop{\longrightarrow}\limits_{x \rightarrow \pm \infty}
\left(\frac{\lambda}{\Lambda}\right)^{1/4}
\left[\frac{\Lambda}{4\left(\Lambda + \varphi^2(x)\right)}\right]^{1/4}
\exp{\left(\frac{1}{2}\ln{\left(\frac{2|S_0(x)|}{\sqrt\lambda}\right)}\right)}
\,.
\end{equation}
The factors of $|S_0(x)|$ remaining in these asymptotic expansions
will all cancel.

For convenience in applying the initial conditions, it is useful to
write
\begin{equation}
I_p(x,x_0) = I_p(x_i,x_0) + I_p(x,x_i)\,,
\end{equation}
where the first term on the right is a complex constant and the second
is real and varying (cf. Eq.~(\ref{eq:Ip_ri})).  After some algebra,
we find that the asymptotic expansion of the general solution in
Eq.~(\ref{eq:gPsi}) can be written as
\begin{eqnarray}
\Psi(x) & \mathop{\longrightarrow}\limits_{x \rightarrow \pm \infty} &
\left[\frac{\Lambda}{4\left(\Lambda + \varphi^2(x)\right)}\right]^{1/4}
\nonumber \\
& &
\times
\left\{
C_\pm
\left(\frac{\varphi +
\sqrt{\Lambda + \varphi^2}}{\sqrt{\Lambda}}\right)^{\pm\eta/2}
\exp\left(\mp\frac{i}{\lambda}I_p(x,x_i)\right)
\right.
\nonumber \\
& &
\left.
+ \;
D_\pm
\left(\frac{\varphi +
\sqrt{\Lambda + \varphi^2}}{\sqrt{\Lambda}}\right)^{\mp\eta/2}
\exp\left(\pm\frac{i}{\lambda}I_p(x,x_i)\right)
\right\}\,.
\label{eq:aPsi}
\end{eqnarray}
In the arguments of the exponentials, all terms of ${\cal O}(\lambda)$
or that vanish as $|x| \rightarrow \infty$ have been dropped.  The
constant coefficients are given by:
\begin{equation}
C_+ =
C \exp\left(-\frac{i}{\lambda}{\rm Re} I_p(x_i,x_0)\right)
\left(A + B e^{-i\nu\pi}\right)\,,
\label{eq:C+}
\end{equation}
\begin{equation}
C_- =
C \exp\left(+\frac{i}{\lambda}{\rm Re} I_p(x_i,x_0)\right)
\left(A e^{-i\nu\pi} + B\right)\,,
\label{eq:C-}
\end{equation}
\begin{equation}
D_+ =
D \exp\left(+\frac{i}{\lambda}{\rm Re} I_p(x_i,x_0)\right)
\left(B e^{-i\nu\pi}\right)\,,
\label{eq:D+}
\end{equation}
\begin{equation}
D_- =
D \exp\left(-\frac{i}{\lambda}{\rm Re} I_p(x_i,x_0)\right)
\left(A e^{-i\nu\pi}\right)\,,
\label{eq:D-}
\end{equation}
and
\begin{equation}
C =
\left(\frac{\lambda}{\Lambda}\right)^{1/4}
\left(\frac{\Omega}{\lambda}\right)^{-i\Omega/4\lambda + \eta/4}
\left(\frac{e^{i\pi/4}}{\sqrt{2}}\right)^\nu
\exp\left(\frac{i\Omega}{4\lambda}\right)\,,
\label{eq:C}
\end{equation}
\begin{equation}
D =
-\frac{\sqrt{2\pi}}{\Gamma(-\nu)}
\left(\frac{\lambda}{\Lambda}\right)^{1/4}
\left(\frac{\Omega}{\lambda}\right)^{+i\Omega/4\lambda - \eta/4}
\left(\frac{e^{-3 i\pi/4}}{\sqrt{2}}\right)^{-\nu -1}
\exp\left(-\frac{i\Omega}{4\lambda}\right)\,.
\label{eq:D}
\end{equation}
Note that the arguments to the $x$-dependent exponentials above are
purely imaginary since $I_p(x,x_i)$ is real (that knowledge will be
convenient when we take the squared modulus of the wave function).



\newpage
\centerline{\bf Figure Captions}

\bigskip
FIG. 1.  The electron neutrino survival probability vs. the
mass-squared difference parameter for two different vacuum mixing
angles.  The solid line is given by the method of this paper.  The
dashed line is the exact (numerical) result.  The dotted line is the
linear Landau-Zener result.  In the top figure, the lines are
indistinguishable. An exponential density with parameters chosen to
approximate the sun was used \cite{Bahcall-dens}.  The region leftward
of the lower left corner of the trough is the nonadiabatic region.

\bigskip
FIG. 2.  The source term (the survival probability, averaged over
detector, minus the survival probability, averaged over both source
and detector) in the electron neutrino survival probability vs. the
mass-squared difference parameter for $\sin^2{2\theta_v} = 0.7$. The
solid line is given by the method of this paper.  The dashed line is
the exact (numerical) result.  The density profile is as in Fig. 1.

\bigskip
FIG. 3.  The source term (the survival probability, averaged over
detector, minus the survival probability, averaged over both source
and detector) in the electron neutrino survival probability vs. the
mass-squared difference parameter for $\sin^2{2\theta_v} = 0.01$.  The
solid line is given by the method of this paper.  The dashed line is
the exact (numerical) result.  The lines are indistinguishable, even
when a zoom is performed in the region of rapid oscillations.  The
density profile is as in Fig. 1.


\newpage

\epsfbox{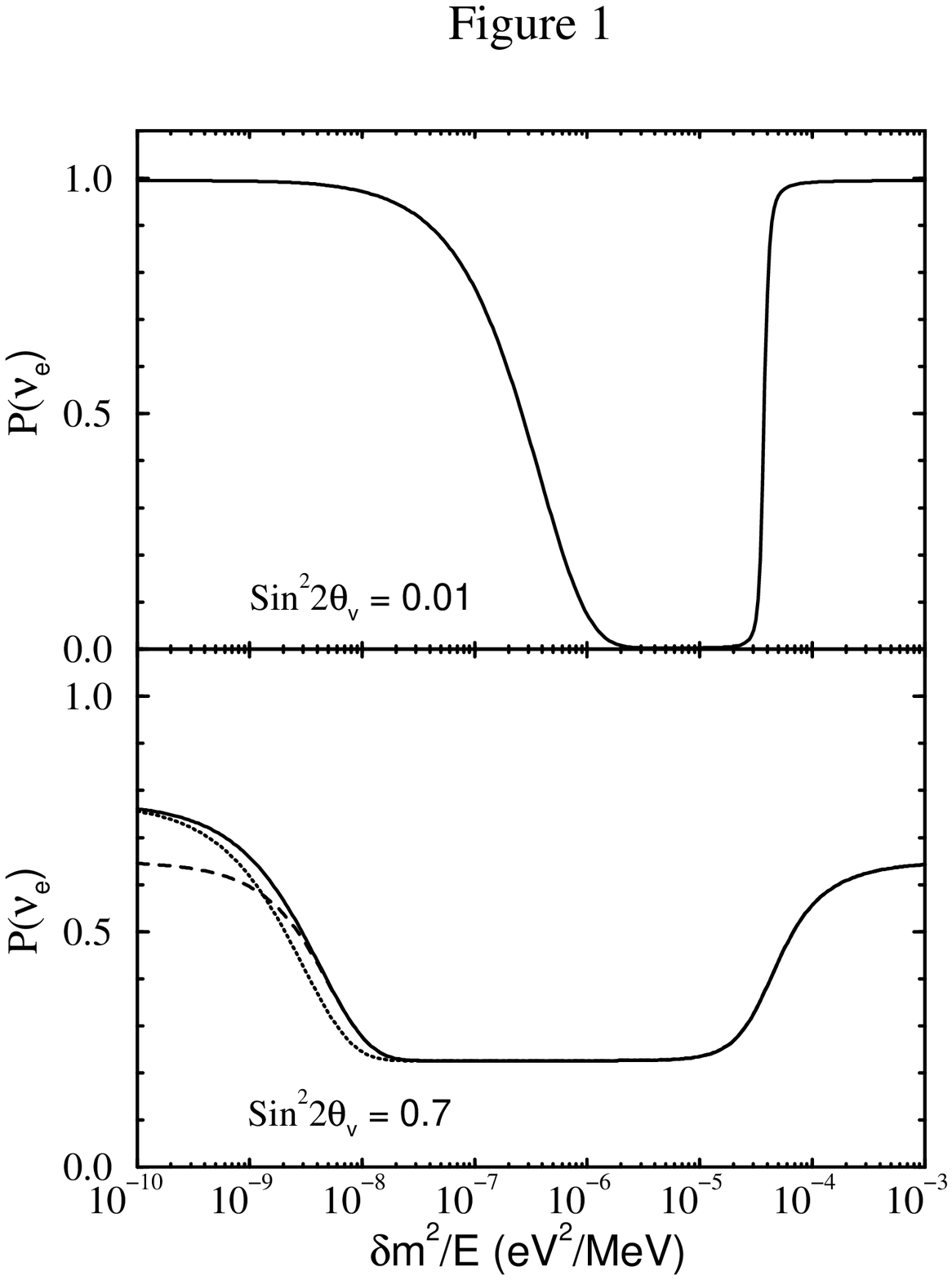}
\epsfxsize=7in \epsfbox{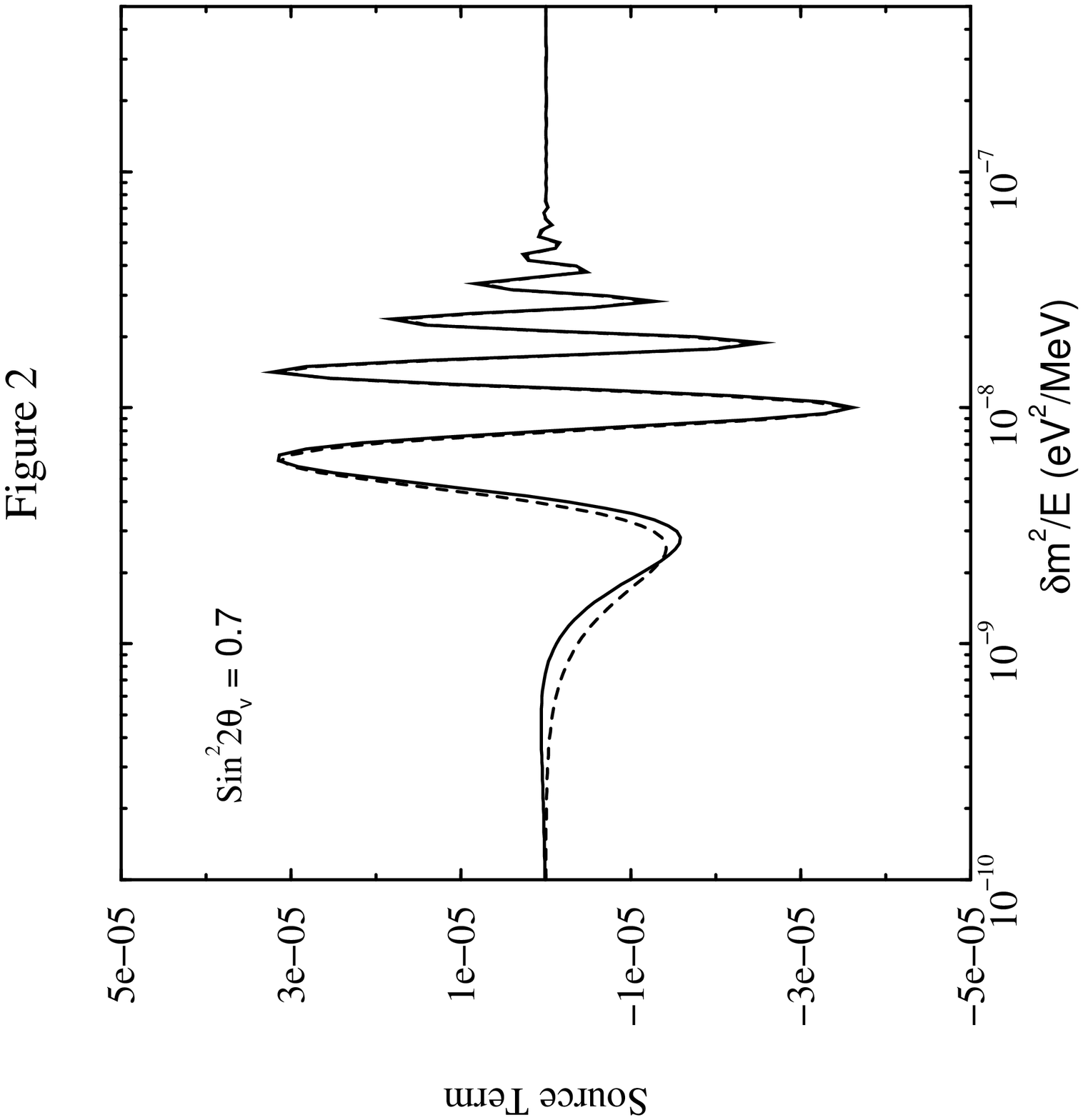}
\epsfxsize=7in \epsfbox{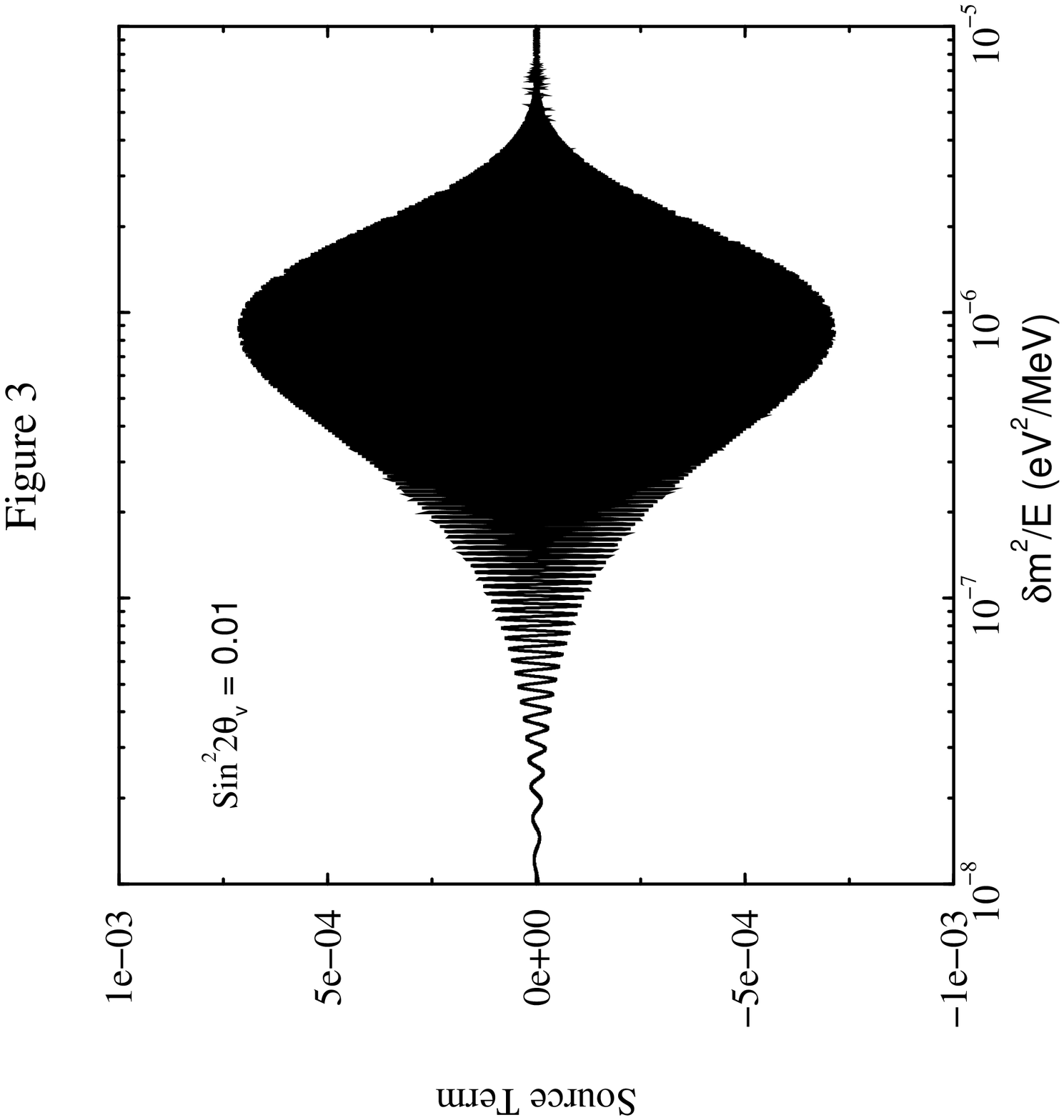}


\end{document}